\documentclass[letterpaper]{JHEP3}

\usepackage{epsfig}

\title{
Noncommutative geometry inspired black holes in higher dimensions at the LHC
}

\author{
Douglas M. Gingrich\footnote{Also at TRIUMF, Vancouver, BC V6T 2A3
Canada.}\\  
Centre for Particle Physics, Department of Physics, University
of Alberta,\\
Edmonton, AB T6G 2G7 Canada\\  
E-mail: \email{gingrich@ualberta.ca}
}

\abstract{
When embedding models of noncommutative geometry inspired black holes
into the peridium of large extra dimensions, it is natural to relate
the noncommutativity scale to the higher-dimensional Planck scale.
If the Planck scale is of the order of a TeV, noncommutative geometry
inspired black holes could become accessible to experiments.
In this paper, we present a detailed phenomenological study of the
production and decay of these black holes at the Large Hadron Collider
(LHC).  
Noncommutative inspired black holes are relatively cold and can be well 
described by the microcanonical ensemble during their entire decay.
One of the main consequences of the model is the existence of a black
hole remnant. 
The mass of the black hole remnant increases with decreasing mass scale
associated with noncommutative and decreasing number of dimensions.
The experimental signatures could be quite different from previous
studies of black holes and remnants at the LHC since the mass of the
remnant could be well above the Planck scale.   
Although the black hole remnant can be very heavy, and perhaps even
charged, it could result in very little activity in the central
detectors of the LHC experiments, when compared to the usual anticipated
black hole signatures. 
If this type of noncommutative inspired black hole can be produced
and detected, it would result in an additional mass threshold above the
Planck scale at which new physics occurs.
}


\preprint{}

\received{\today}

\begin{document}

\section{Introduction\label{sec1}}

The concept of the quantization of spacetime has existed since the
development of quantum mechanics and was consider as a way to help
regulate the short-distances behaviour of point
interactions~\cite{Snyder}. 
The correspondence between geometry and noncommutative algebras was
developed formally in the theory of noncommutative geometry~\cite{Connes}. 
String theory and M-theory have renewed interest in the quantization
of spacetime in terms of the generalization of quantum field theory to
noncommutative coordinates~\cite{Witten96,Seiberg99} (See
Ref.~\cite{Douglas} for a review.).   
In noncommutative quantum field theory, the spacetime coordinates can be
thought of as operators, and the commutator of two coordinates
$\hat{x}^A$ and $\hat{x}^B$ in $D$-dimensions is nonzero: 

\begin{equation} \label{eq1}
\left[ \hat{x}^A, \hat{x}^B \right] = i\theta^{AB} \equiv i
\frac{\epsilon^{AB}}{\Lambda_\mathrm{NC}^2}\, , 
\end{equation}

\noindent
where $\theta^{AB}$ is an real antisymmetric $D\times D$ matrix.
For convenience, we have separated the mass scale $\Lambda_\mathrm{NC}$
associated with noncommutative from the dimensionless matrix structure
$\epsilon^{AB}$ of $\theta^{AB}$.
If $\Lambda_\mathrm{NC}^{-2}$ is the average magnitude of the elements
in $\theta^{AB}$, we assume the elements of $\epsilon^{AB}$ are of 
$\mathcal{O}(1)$.

We would ideally like to construct the noncommutative equivalent of
general relativity. 
Since noncommutativity is an intrinsic property of the manifold itself, 
rather than a superimposed geometric structure due to the fields,
the noncommutative approach could lead to a deeper level of
understanding of gravity.  
It may even contain some of the underlying features of a full theory of
quantum gravity. 
Unfortunately, progress in this area is not yet mature enough to allow
phenomenological studies.

Some progress in understanding the effects of noncommutative gravity can
be made by formulating a model in which general relativity is its usual
commutative form but the noncommutative geometry leads to a smearing of
matter distributions on a length scale of
$\mathcal{O}(1/\Lambda_\mathrm{NC})$.    
The smearing is viewed as due to the intrinsic uncertainty embodied in
the coordinate commutator of Eq.~(\ref{eq1}).
This effective approach could be considered as an improvement to
semiclassical gravity.

Models for noncommutative geometry inspired chargeless, nonrotating
black holes that use this idea were developed by Nicolini, Smailagic, and
Spallucci~\cite{Nicolini06}.  
The model was extended to the case of charge in four
dimensions~\cite{Ansoldi06}, generalized to higher dimensions by
Rizzo~\cite{Rizzo06b}, and then to charge in higher
dimensions~\cite{Spallucci09}.
A review of these developments can be found in Ref.~\cite{Nicolini09}.

Within this effective theory, the point-like sources in the
energy-momentum tensor, which are normally represented by
delta functions of position, are now replaced by smeared matter
distributions of width $\sqrt{\theta}$.  
The smearing is usually taken as gaussian:

\begin{equation} \label{eq2}
\rho = \frac{m}{(4\pi\theta)^{(n+3)/2}} e^{-r^2/(4\theta)}\, ,
\end{equation}

\noindent
where $r$ is the radial distance in $D-1 = n + 3$ space dimensions
from the most probable value of the mass $m$.
The width of the gaussian smearing should be related to the noncommutativity
scale and for simplicity we take $\sqrt{\theta} = 1/\Lambda_\mathrm{NC}$.
The mass is only part of the total mass-energy of the system. 
The energy stored in long-range fields will also contribute to 
the total mass-energy sources of the gravitational field.
Throughout this paper, we will ignore the effects of long-range fields
such as those due to electric charge and colour.
To treat these properly would require a theory of noncommutative
electrodynamics and chromodynamics.
Progress has been made in incorporating electric
charge~\cite{Ansoldi06,Spallucci09}, but we leave the study of its
implications for the Large Hadron Collider (LHC) to future work. 

To observe the effects of noncommutative, one can postulate that
$\Lambda_\mathrm{NC}$ is a new scale greater than about a TeV.
However, we will take the approach that the noncommutativity scale is
related to the reduced Planck scale in models of low-scale  
gravity~\cite{Arkani98,Antoniadis98,Randall99a,Randall99b,Dvali07a,Dvali07b}.
In particular, we consider the ADD~\cite{Arkani98,Antoniadis98} paradigm
of low-scale gravity, and take $\Lambda_\mathrm{NC} \sim M_D$, where
$M_D$ is the $D$-dimensional Planck scale defined by the
PDG~\cite{PDG}\footnote{We use the $M_D$ definition of the Planck scale
since this is usually the quantity experiments set limits on.   
Ref.~\cite{Rizzo06b} used the reduced higher-dimensional Planck scale
$M_*^{n+2} \equiv \bar{M}_D^{n+2} = M_D^{n+2}/(2\pi)^n$.}.

By employing a noncommutative ADD model, we can determine the
gravitational radius of black holes.
From this, the phenomenology of black hole production and decay in
particle collisions follows in the usual
way~\cite{Argyres98,Banks99,Dimopoulos01,Giddings01a}.
By considering experimental limits on the Planck scale and the energy
reach of the LHC, we can restrict the parameter space.
This allows us to make predictions that could guide searches by
experiments at the LHC.  
With the startup of the LHC, such phenomenological studies are important
and of value. 

An outline of this paper is as follows.
In Sec.~\ref{sec2}, we restrict the parameter space to that accessible
at LHC energies and calculate the proton-proton production cross section
for noncommutative inspired black holes.
This shows that black holes with minium mass above the Planck scale
could be produced at the LHC.
The decay, temperature, and existence of a long-lived remnant are
discussed in Sec.~\ref{sec3}.
Since the black holes are relatively cold, we argue that it is valid to
consider Hawking evaporation all the way down to the remnant mass, if it
is above the Planck scale. 
In Sec.~\ref{sec4}, we present a simulation of noncommutative inspired
black hole production and decay at the LHC.
Based on the signatures, we caution that current search scenarios for 
black holes by the LHC experiments may miss noncommutative inspire black
holes with remnant mass above the Planck scale.  
We conclude in Sec.~\ref{sec5} with a discussion of some of the benefits
of a noncommutative geometry view of gravity over the semiclassical
approach, and point out that a second threshold in the measured
inclusive cross section at the LHC may help reveal noncommutative
geometry effects.

\section{Black Hole Production\label{sec2}}

The production cross section for black holes in particle collisions is
estimated by knowing the gravitational radius.
We consider spherically symmetric, nonrotating black holes with no local
charges in $D=n+4$ dimensions.
If the gravitational radius $r_g$ and $\sqrt{\theta}$ are much
smaller than the compactification scale of the extra dimensions, we will
not be sensitive to the compactification moduli.  
This will be a valid assumption for the range of extra dimensions we
consider. 
The $D$-dimensional spherically symmetric and static solutions for a
mass $m$ have been previously obtained by Rizzo~\cite{Rizzo06b}:

\begin{equation} \label{eq3}
\frac{m}{M_D} =
\frac{k_n}{P\left(\frac{n+3}{2},\frac{r_g^2}{4\theta}\right)} \left( r_g
M_D \right)^{n+1}\, , 
\end{equation}

\noindent
where 

\begin{equation} \label{eq4}
k_n = \frac{n+2}{2^n \pi^{(n-3)/2} \Gamma\left(\frac{n+3}{2}\right)}
\end{equation}

\noindent
and

\begin{equation} \label{eq5}
P\left(\frac{n+3}{2},\frac{r_g^2}{4\theta}\right) =
\frac{1}{\Gamma\left(\frac{n+3}{2}\right)} \int_0^{r_g^2/(4\theta)} dt\,
e^{-t} t^{(n+3)/2-1}\, .  
\end{equation}

\noindent
The constant $k_n$ depends only on $n$ and the definition of the
Planck scale.\footnote{Our expression for the constant $k_n$ is
related to $c_n$ of Ref.~\cite{Rizzo06b} by $c_n = (2\pi)^n k_n$
since we have used $M_D$ rather than $\bar{M}_D$.  
By redefining this constant, the relationship between $m$ and $r_g$
in Eq.~(\ref{eq3}) is the same as in Ref.~\cite{Rizzo06b}.} 
The function $P$ is the regularized incomplete gamma function from
below.  
Using a different smearing function would simply replace the function
$P$ in Eq.~(\ref{eq3}) with an alternative function. 
In Eq.~(\ref{eq3}), it is not possible to write $r_g$ as a
function of $m$ as done in the commutative case.

Since $n$ is an integer, the incomplete gamma function takes on
restricted functional forms and we use a simpler notation similar to
Ref.~\cite{Rizzo06b}: $F_a(q) \equiv
P\left[(n+3)/2,r_g^2/(4\theta)\right]$, where $a = (n+3)/2$ and $q =
r_g^2/(4\theta)$. 
Although closed forms for $F_a$ are known for integer $a$, we will
always solve $F_a$ numerically using either series or continued
fraction approximations, since the approximations are quite good. 

The new features of noncommutative inspired black holes as compared to
the commutative case are embodied in the properties of the incomplete
gamma function.
Some of the properties of $F_a(q)$ are:
$F_a(q) \to 1$ as $q \to \infty$ and $F_a(q) \to 0$ as $q \to 0$, for
all values of $a$.
More precisely $F_a(q) \sim q^a$ as $q\to 0$.  
Thus, for $r_g \gg \sqrt{\theta}$, $F_a(q)\to 1$, and we obtain the
standard commutative result, which can be inverted to obtain the
gravitational radius as a function of mass. 
Also, we reproduce the noncommutative four dimensional
result~\cite{Nicolini06} when $n\to 0$. 
For LHC energies and $n$ in the range 1 to 7, neither of the above
limits are particularly accurate approximations of $F_a(q)$.
Thus, we will always take $F_a(q)$ fully into account in calculations.

Ref.~\cite{Rizzo06b} has shown that the detailed structure of the
smearing distribution is not important as long as we do not probe
distances smaller than $\sim 1/\Lambda_\mathrm{NC}$.
This helps justify our earlier choice of $\sqrt{\theta} =
1/\Lambda_\mathrm{NC}$, since we can expect that different choices for 
the smearing distribution, parameterized by the length scale
$\sqrt{\theta}$,  might change this relationship by only
$\mathcal{O}(1)$.  
To be independent of the choice of distribution, we will only consider
$r_g > 1/\Lambda_\mathrm{NC}$. 
Besides being sensitive to the details of the smearing distribution at
$r_g \sim 1/\Lambda_\mathrm{NC}$, we might also expect the onset of
corrections due to quantum gravity. 

By examining Eq.~(\ref{eq3}) and the properties of the incomplete gamma
function, we see that as $r_g \to \infty$, $F_a \to 1$, and hence $m
\to \infty$. 
Also as $r_g \to 0$, $m \sim r_g^{-2} \to \infty$.
So a minimum value of $m = m_\mathrm{min}$ exists for some finite
positive $r_g = (r_g)_\mathrm{min}$. 
Because there is a minimum mass, there are masses below which the black
hole will not form, and above the minimum mass the gravitational radius
is double valued.
However, it is well behaved and never vanishes.
As the mass increases, the inner gravitational radius shrinks to zero,
while the outer gravitational radius approaches the noncommutative
value. 
These observations were first made in Ref.~\cite{Nicolini06,Rizzo06b}.
For reasons related to positive temperature given later, only the outer
gravitational radius $r_g\ge (r_g)_\mathrm{min}$ is relevant to us.  

We now show that the experimental lower bounds on $M_D$ and the maximum
energy of the LHC will restrict the values of $\sqrt{\theta}$ that can
be probed by experiments at the LHC.   
We do not expect black holes to form for masses much less than $M_D$. 
This give a lower bound on $m_\mathrm{min}$.
We will consider only the hard limits on the Planck scale from
measurements of deviations from Newton's gravity and limits set by
accelerator experiments: 
$M_D > 4$~TeV for $n = 2$~\cite{PDG,Kapner07}, 
$M_D > 1.2$~TeV for $n = 3$, 
$M_D > 0.94$~TeV for $n = 4$~\cite{DELPHI,L304,LEP}, 
$M_D > 0.86$~TeV for $n = 5$, 
$M_D > 0.83$~TeV for $n = 6$ ~\cite{CDF,D0}, and
$M_D > 0.80$~TeV for $n = 7$~\cite{D0}.
The astrophysical and cosmological limits are higher, particularly for
two or three extra dimensions.
However, they are based on a number of assumptions so the results are
only order of magnitude estimates.
Throughout this paper, we will restrict our considerations to $2 \le n
\le 7$, although higher dimensions are not excluded.
We explicitly consider only ADD-type black holes.
However, there is a range of mass scales for which almost flat
five-dimensional space is an applicable metric for Randall-Sundrum
type-1 black holes.
With suitable relationships between the parameters of the models, $n=1$
could also be considered.  
The maximum mass of the black hole is likely to be limited by the
statistics of the maximum parton energies in a proton-proton
interaction, but in no case can it be larger than the proton-proton
centre-of-mass energy. 
Thus, we will only be interested in the case where the minimum black
hole mass is below the LHC maximum energy of 14~TeV and above the
experimental lower bound on the Planck scale.

Follow Ref~\cite{Rizzo06b}, we use Eq.~(\ref{eq3}) to calculate
$\partial m/\partial r_g = 0$ for fixed $\sqrt{\theta}$ to obtain 

\begin{equation} \label{eq6}
F_a(q_0) -
\frac{2q_0^a e^{-q_0}}{(n+1)\Gamma(a)} = 0\, , 
\end{equation}

\noindent
where $q_0$ is the root of the resulting equation, which gives the
gravitational radius at which the minimum mass occurs.
Equation~(\ref{eq6}) does not depend on $k_n$, i.e.\ the definition of
the Planck scale.
We can now solve Eq.~(\ref{eq6}) for $q_0$, for each value of $a$ (or $n$).
For each value of $q_0$, we calculate $\sqrt{q_0}/2 =
(r_g)_\mathrm{min}/\sqrt{\theta}$.
The values of $(r_g)_\mathrm{min}/\sqrt{\theta}$ depend only on the
number of dimensions. 
From the values of $(r_g)_\mathrm{min}/\sqrt{\theta}$, we calculated
$(r_g)_\mathrm{min}$ and thus $m_\mathrm{min}$, for a given value of
$\sqrt{\theta}$ and $n$.  

We obtain a valid range of $\sqrt{\theta} M_D$ for each number of
dimensions by restricting the minimum black hole mass at the LHC to be
in the range $1 < m_\mathrm{min}/M_D < 14~\mathrm{TeV}/M_D$, as
discussed above.
Increasing $M_D$ restricts the range of masses.
The results are given in Table~\ref{tab1}.
We see that the gravitational radius at the minimum mass is not
particularly sensitive to the number of dimensions, differing by less
than 8\% over the range $3 \le n \le 7$. 
The ranges of $\sqrt{\theta}M_D$ are restricted to intervals of about
0.3, but there is no single value of $\sqrt{\theta} M_D$ that lies
in the allowed range for all number of dimensions\footnote{In
Ref.~\cite{Rizzo06b} a single region of $\sqrt{\theta}$ was obtained
because of the different definition of the Planck scale. 
Effectively $\sqrt{\theta} \to (2\pi)^{n/(n+2)} \sqrt{\theta}$, since
Eq.~(\ref{eq6}) does not depend on $k_n$.}.
If $\sqrt{\theta} \ll 1/M_D$, the minimum black hole mass becomes zero
as we approach the commutative limit.
If $\sqrt{\theta} \gg 1/M_D$, the black hole minimum mass is beyond the
energy reach of the LHC, unless the number of dimensions is very large.

\TABLE[h]{
\centering
\begin{tabular}{|c|c|c|c|c|}\hline
$n$ & $(r_g)_\mathrm{min}/\sqrt{\theta}$ &
$(m_\mathrm{min}/M_D) (\sqrt{\theta} M_D)^{-(n+1)}$ &
$\sqrt{\theta}_\mathrm{min}M_D$ &
$\sqrt{\theta}_\mathrm{max}M_D$\\\hline
2 & 2.51 & 65.2 & 0.248 & 0.377\\
3 & 2.41 & 58.8 & 0.361 & 0.667\\
4 & 2.34 & 48.6 & 0.460 & 0.789\\
5 & 2.29 & 37.9 & 0.546 & 0.869\\
6 & 2.26 & 28.2 & 0.621 & 0.929\\
7 & 2.23 & 20.3 & 0.686 & 0.982\\
\hline
\end{tabular}
\caption{Values of minimum gravitational radius $(r_g)_\mathrm{min}$ in
units of $\sqrt{\theta}$ and minimum mass $m_\mathrm{min}$ in units of
$M_D (\sqrt{\theta} M_D)^{n+1}$.  
The last two columns show the range of $\sqrt{\theta}$ in units of
$1/M_D$ that can be probed at the Large Hadron Collider.}
\label{tab1}
}

So as not to be sensitive to the details of the smearing distribution, we
want to be a few standard deviations away from the mean of the gaussian
distribution, that is, $r_g > \sqrt{2\theta}$. 
In Table~\ref{tab1}, we see that $(r_g)_\mathrm{min} > 2.2
\sqrt{\theta}$ (1.6 standard deviations) provided $n \le 7$.
Thus, we are probing, at most, 10\%/2 = 5\% of the distribution when the
gravitational radius is its minimum value.
At $(r_g)_\mathrm{min}$, we might be sensitive to the type of
smearing distribution but this is probably only an $\mathcal{O}(1)$
effect. 

The model of noncommutative geometry inspired black holes in higher
dimensions has three unknown parameters: $n$, $M_D$, and
$\sqrt{\theta}$. 
However, in this paper, we are only interested the parameter space that
can be probed by the LHC.
Thus by choosing the relationship between $M_D$ and $\sqrt{\theta}$ to
be consistent with experimental bounds and accessible to the LHC, we
restrict the parameter space.
What is important is not the exact numerical relationship between these
constants but that our choice does not exclude a type of black hole
signature that could occur within the model. 

\TABLE[hr]{
\centering
\begin{tabular}{|c|c|c|}\hline
$n$ & $(r_g)_\mathrm{min}M_D$ & $m_\mathrm{min}/M_D$\\\hline
2 & 1.51 & 14.09\\
3 & 1.45 &  7.62\\
4 & 1.40 &  3.78\\
5 & 1.38 &  1.77\\
6 & 1.35 &  0.79\\
7 & 1.34 &  0.34\\
\hline
\end{tabular}
\caption{Minimum gravitational radius and minimum mass for
$\sqrt{\theta} M_D = 0.6$.} 
\label{tab2}
}

The values for the minimum gravitational radius and mass for the choice
of $\sqrt{\theta} M_D = 0.6$ are shown in Table~\ref{tab2}.
The minimum gravitational radius just scales linearly with
$\sqrt{\theta}$.
For our choice of noncommutative scale, the $n=2$ case is just out 
of the energy reach of the LHC.
This is not too biased of a choice since the Planck scale lower limit
for two extra dimensions is quite high and the chances of even a
commutative version of this black hole being produced at the LHC is
low. 
Table~\ref{tab2} shows that we can examine three different types of
noncommutative black holes by choosing different number of dimensions. 
If $n \le 2$, the minimum mass of the black hole is above the LHC energy
reach; 
if $n \ge 6$, the minimum mass of the black hole is below the Planck
scale;
and for $3 \le n \le 5$ the minimum mass is within the LHC energy reach.   
Considering a different relationship between $\sqrt{\theta}$ and $M_D$
within the allowed range in Table~\ref{tab1}, would shift the values of
the minimum black hole masses but would not change the phenomenology of
the three types of noncommutative black holes. 
The results are expressed in units of $M_D$, and thus if $M_D$ is further
constrained by future experiments, Table~\ref{tab1} and \ref{tab2} can
still be used.  

We need to decide how to handle the case when the minimum gravitational
mass is below the Planck scale.
We might imagine in a UV complete theory of quantum gravity, that the
threshold for black hole production in this case is close to the Planck
scale.  
As a working hypothesis, it is not unreasonable to take the minimum mass
of the black hole to be the maximum of $M_D$ or $m_\mathrm{min}$. 

A feature of black hole production in particle collisions is that the
cross section is essentially the horizon area of the forming black hole
and grows with the centre-of-mass energy of the colliding particles as
some power. 
Assuming the black hole production cross section is equal to the
geometrical area, we use $\hat{\sigma} = \pi r_g^2$ for the hard parton
cross section.
This neglects parton charge, colour, spin, and finite size (See
Ref~\cite{Gingrich06a} for a review on these developments in the
commutative ADD case.). 
Figure~\ref{fig1} shows the parton cross section for $\sqrt{\theta} M_D
= 0.6$ and different number of dimensions.
The minimum mass for the $n=2$ case occurs just above the LHC maximum
energy and thus does not appear in Fig.~\ref{fig1}.
Also shown for comparison are the usual commutative cross sections.
The most significant differences occur for low number of dimensions and
at minimum black hole masses.

\FIGURE[ht]{
\epsfig{file=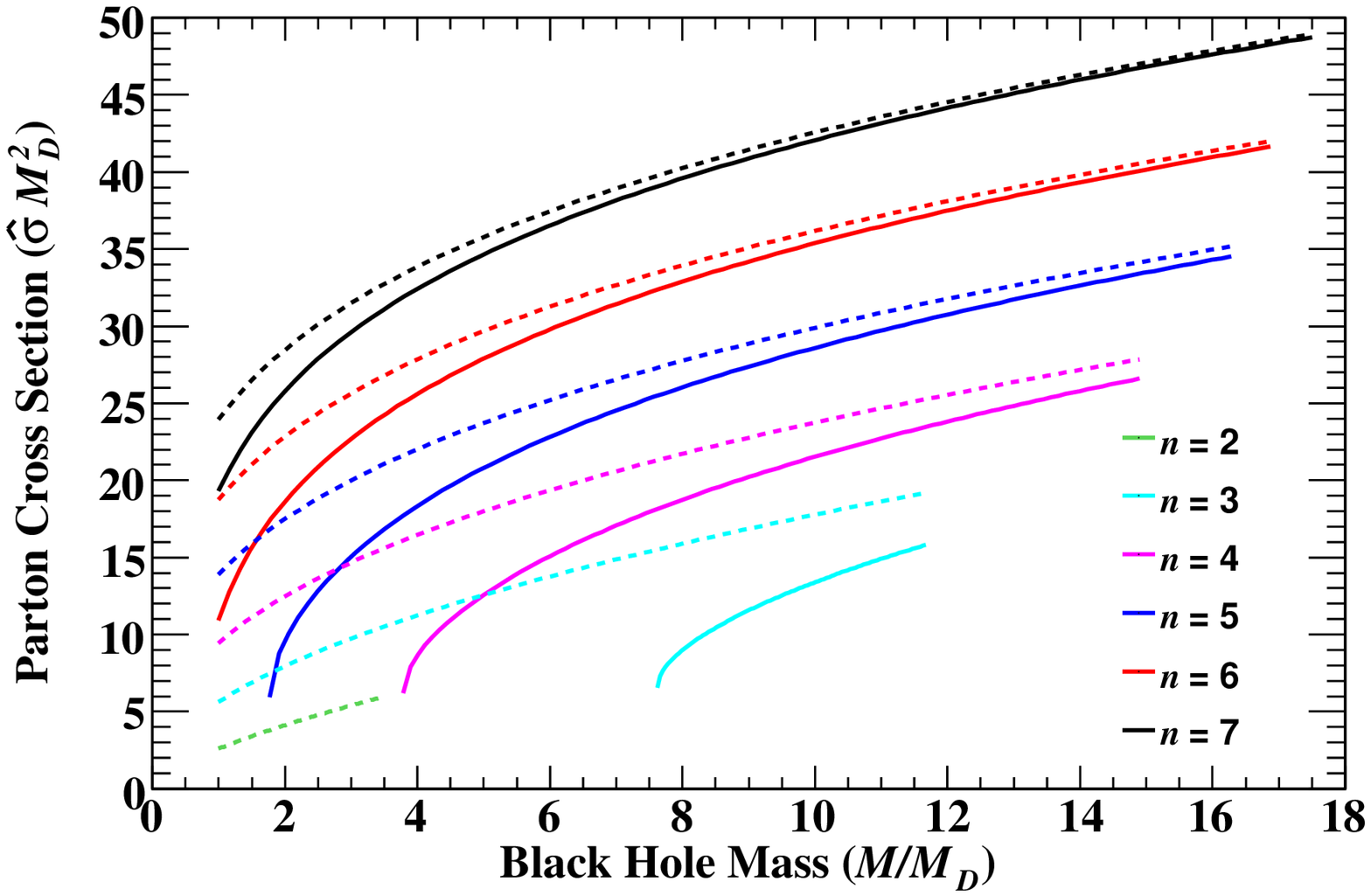,width=12cm}
\caption{Parton cross section versus black hole mass for $\sqrt{\theta}
M_D = 0.6$.  
The solid lines are for noncommutative inspired black holes and 
the dashed lines for the commutative ADD black holes.
The number of dimensions decreases from top to bottom.}
\label{fig1}
}

Only a fraction of the total centre-of-mass energy $\sqrt{s}$ in a
proton-proton collision is available in the hard scattering process.
We define $s x_a x_b \equiv s x_\mathrm{min} \equiv \hat{s}$, where
$x_a$ and $x_b$ are the fractional energies of the two partons relative
to the proton energies.
The full particle-level cross section $\sigma$ is given by

\begin{equation} \label{eq7}
\sigma = \sum_{a,b} \int^1_{M^2/s} d
x_\mathrm{min} \int^1_{x_\mathrm{min}} \frac{dx}{x} f_a \left(
\frac{x_\mathrm{min}}{x} \right) f_b(x) \pi r_g^2\, , 
\end{equation}

\noindent
where $a$ and $b$ are the parton types in the two protons, and $f_a$ and
$f_b$ are the parton distribution functions (PDFs) for the proton.
The sum is over all the possible quark, antiquark, and gluon pairings. 
The parton distributions fall rapidly at high fractional energies, and
so the particle-level cross section falls at high energies.

When calculating the particle-level cross sections, we integrate
Eq.~(\ref{eq7}) down to the experimental bound on the Planck scale or
minimum mass, whichever is higher, for each number of extra dimensions.
The MRST LO* parton distribution functions~\cite{Sherstnev} were used
with a QCD scale of $Q = 1/r_g$.
The cross sections are sensitive to the choice of PDFs and QCD scale.
A difference of about 7\% is observed for different PDFs and if $Q = M$
is used for the QCD scale, the cross sections are about 13\%
lower~\cite{Gingrich09b}.  
To obtain numerical results, we have used the parameters shown in
Table~\ref{tab2}.    
You may assume these parameters were used in calculations unless told
otherwise. 

Figure~\ref{fig2} shows the total proton-proton cross section versus
lower mass threshold for $\sqrt{\theta} M_D = 0.6$ and different number
of dimensions.   
Also shown for comparison are the usual commutative cross sections.
We notice that differences between the noncommutative and commutative 
cross sections are minimal, on a logarithmic scale, provided the minimum
mass is above threshold. 
The biggest differences are near the minimum mass for small number of
extra dimensions.
Thus noncommutative effects in the cross section will probably only be 
observable if the minimum mass is above the Planck scale.

\FIGURE[ht]{
\epsfig{file=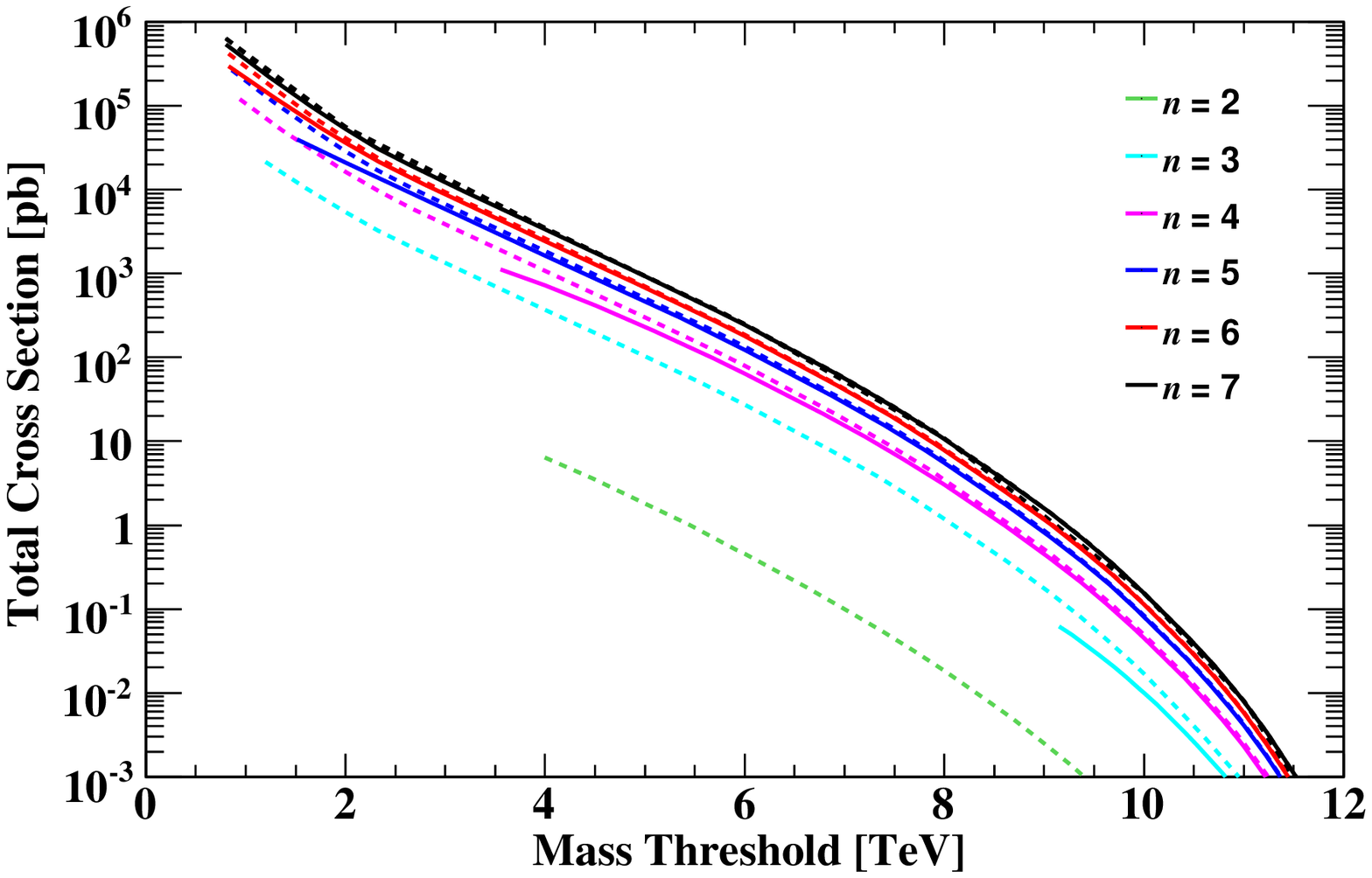,width=12cm}
\caption{Proton cross section versus a threshold mass for $\sqrt{\theta}
M_D = 0.6$. 
The solid lines are for noncommutative inspired black holes and the
dashed lines for  the commutative ADD black holes.
The number of dimensions decrease from top to bottom.}
\label{fig2}
}

Mechanisms that we can think of to lower the cross section, such as
initial radiation before the black hole is formed, should apply equally 
well to noncommutative and commutative black holes.
The validity of the cross sections at low threshold masses are
questionable.
The cross sections are based on classical arguments, which are most
likely only valid when the threshold mass is significantly above the
Planck scale.
For the noncommutative case and $3 \le n \le 5$, the threshold masses
are above the Planck scales and hence may be valid all the way down to
the minimum masses. 
Particularly in the case of $n=3$ (and $\sqrt{\theta}M_D = 0.6$), we 
might expect the cross section to be valid over the entire mass range
except at the minimum mass, where quantum corrections may be
anticipated.    

The important point is not so much the value of the cross sections,
besides that they are high, but that there is little difference between
the noncommutative and commutative cross sections once we are above the
lower-mass threshold. 
It will be the responsibility of the experiments to measure or set upper 
limits on the proton-proton cross section.
In the case of low-scale gravity at about a TeV, an increase in the
inclusive cross section versus mass should be an indication of new
physics at a mass threshold approximately equal to the scale of this new
physics.  
If a second increase in cross section occurs at a higher energy, this
may be an indication of noncommutative inspired black holes with minimum
black hole mass near this second threshold.
Thus, although there is only one new physics scale, the experimental
signatures may indicate more than one new production threshold.

We remind the reader that we have not taken electromagnetic charge into
account.
If the electromagnetic field is allowed to spread out into the
bulk, the energy of the electrostatic field increases the total
mass-energy~\cite{Spallucci09}.  
Thus, the minimal value of the black hole mass can be obtained by
studying the neutral case, as we have done.  

\section{Black Hole Decays} \label{sec3}

We now consider noncommutative inspired black hole decays.
Of primary importance is the fact that noncommutative black holes have a
minimum mass, which is usually taken to be the mass of a remnant.
Many models predict remnants. 
A few references with relevance to the LHC are models of higher
curvature invariants in the action~\cite{Rizzo05a,Rizzo05b,Rizzo06a},  
models of minimum length scale~\cite{Hossenfelder04}, 
models in which Newton's constant is take to be a running
parameter~\cite{Bonanno00}, 
loop quantum gravity~\cite{Bojowald05}, 
tunnelling~\cite{Xiang}, 
string gravity~\cite{Alexeyev}, 
a general uncertainty principle~\cite{Adler01,Cavaglia03b},
and from generic considerations~\cite{Hossenfelder03a,Koch05,Hossenfelder05}.  
See Ref.~\cite{Casher} for a review, and Ref.~\cite{Susskin95} for
arguments against remnants.
Many of these models predict the remnant mass is at the Planck scale.
In the noncommutative case for low number of extra dimensions, this may 
not be true and remnants can have significant mass above the Planck
scale. 
Since most black holes in proton-proton collisions would be produced
near threshold, the black hole in remnant models will predominantly be
produced just above the mass of the remnant, allowing for vary little
energy in the decay.  

In this discussion, we ignore any possible balding phase in which the
higher gravitational moments are shed before, during, or after black
hole production.  
Since we are not considering local charges in the metric, we ignored a
possible initial sudden loss of Abelian or non-Abelian hair, and a
Schwinger pair production mechanism for decay.  
Since we have neglected spin, we do not consider a possible spin-down
phase. 
We will restrict our decays to Hawking evaporation.
Black hole accretion is also not considered since it is typically not
important for small black holes.

Given the above limitations, we expect the decay to depend on the
temperature of the black hole.  
The temperature of noncommutative inspired black holes is given
by~\cite{Rizzo06b} 

\begin{equation} \label{eq8}
T = \frac{n+1}{4\pi r_g M_D} \left[ 1 -
\frac{2q^{(n+3)/2}e^{-q}}{F_n(q)(n+1)\Gamma\left(\frac{n+3}{2}\right)}
\right]\, . 
\end{equation}

\noindent
The quantity in square brackets modifies the usual higher-dimensional
commutative form.
In addition, since $r_g$ has been modified, it also leads to temperature 
differences\footnote{The temperature does not depend on $k_n$ so we
expect small differences with respect to Ref.~\cite{Rizzo06b}.}. 

Figure~\ref{fig3} shows the temperature versus mass for $\sqrt{\theta}
M_D = 0.6$.
We notice that the temperature vanishes at the minimum mass.
These observations were first made in Ref.~\cite{Nicolini06,Rizzo06b}.
Also shown for comparison are the usual Hawking temperatures.
At LHC energies, the noncommutative temperature is significantly
different from the Hawking temperature for most masses.
Only for $n \ge 5$ is the mass at the maximum temperature within the LHC
energy range.
The temperature increases with increasing number of dimensions, and for
$n=7$ is never greater than about 130~GeV.
Thus noncommutative inspired black holes are always ``cold'' if produced
at the LHC. 
Since the inner gravitational radius corresponds to negative temperatures,
we have only consider the outer radius.

\FIGURE[ht]{
\epsfig{file=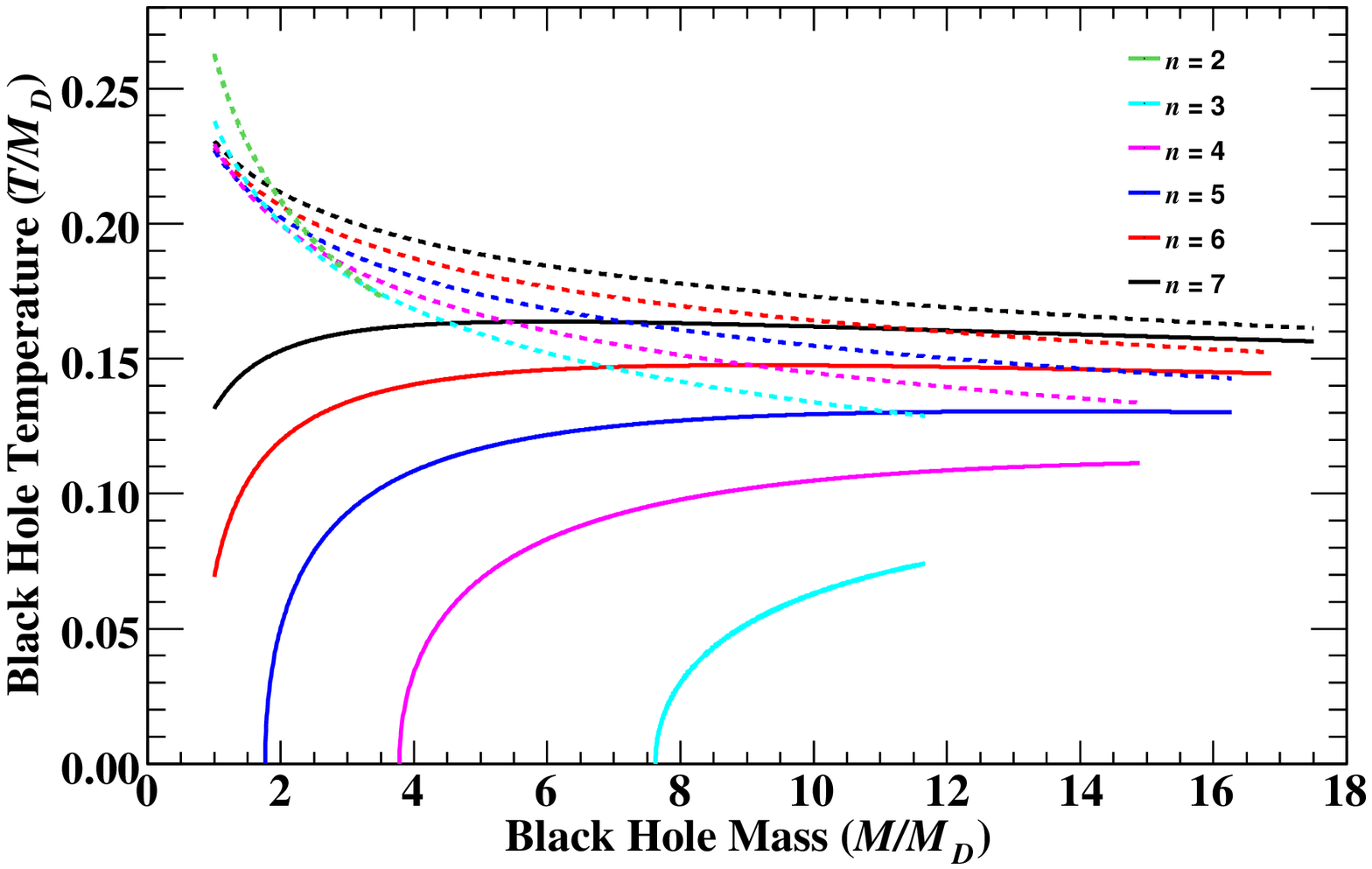,width=12cm}
\caption{Temperature versus black hole mass for $\sqrt{\theta} = 0.6$. 
The solid lines are for noncommutative inspired black holes and the
dashed lines for the commutative ADD black holes.
The number of dimensions decreases from top to bottom.}
\label{fig3}
}

Hawking evaporation is usual considered in the context of the canonical
ensemble~\cite{Gingrich07b}.
The canonical ensemble does not apply when the mass is near the Planck
scale or when the emitted particles carry an energy comparable to the
black hole mass itself, due to the back-reaction of the emitted particles
on the properties of the black hole.
In the commutative case, one expects significant back-reaction effects
during the terminal stage of evaporation because of the increase in
temperature at low masses. 
However, noncommutative black holes are much cooler than commutative
black holes.
The energy of Standard Model particles emitted from the black hole will
always have energy much less than the black hole mass, and thus the back
reaction is never significant. 
The back reaction could be significant near the maximum temperature but
this maximum is not very high.
The typical temperature of such black holes ($\lesssim 130$~GeV)
always remains smaller than their mass.
The microcanonical corrections can thus be neglected.  

We can safely treat the decay according to the usual canonical decay
formalism.
Since the temperature decreases near the minimum mass, we employ Hawking
evaporation during the full decay and do not need to invoke a
terminal-decay phase near the end of the decay.  
Thus no unphysical production or decay thresholds are required for
noncommutative black holes with minimum mass above the Planck scale.

Since the temperature is small near the minimum mass, we might expect
that near the end of the decay a large number of soft zero mass
particles can be emitted. 
This is in contrast to the Hawking case where the probability of
emitting a particle of half the black hole mass is quite probable near
the end of the Hawking evaporation stage.
Thus in addition to the presence of a remnant, the kinematics of the
other particles in the decay will be significantly different for
noncommutative black holes.

The decay times have been studied in Ref.~\cite{Casadio08}.
The black hole decays very quickly down to the remnant and then takes 
an infinite time to decay near the remnant.
The black hole would decay to the remnant within the ATLAS and CMS
detectors.
At a temperature of $\sim 10^{-10}$~MeV, the black hole would be in
thermal equilibrium with the cosmic microwave background.
It has also been shown~\cite{Rizzo06b} that the heat capacity is zero
at the minimum mass,  i.e.\ there will be no further change of the
black hole mass with its temperature. 
The radiation process ends with a finite black hole mass of near zero
temperature.

We will take the black hole remnant to be stable.
Our decision is based purely on the classical arguments of a minimum
mass, zero temperature, and zero heat capacity.
It is unknown if quantum effects can destabilize the remnant.
The final state of black hole decay requires quantum gravity corrections, 
which the usual semiclassical model is unable to provide.
In the noncommutative model, one expects that the later stage of
evaporation is determined by spacetime fluctuations of the manifold, when 
the radius of the black hole horizon becomes comparable with
$\sqrt{\theta}$.   
Since this is the quantum nature of the noncommutative geometry,
we might expect it to mimic some of the effects of quantum gravity.  
Indeed, noncommutative effects might be a good approximation to quantum
gravity, when the remnant mass is above the Planck
scale. 

We remind the reader that we have not taken charge properly into account.
If the electromagnetic field is allowed to spread out into the bulk,
the effect of charge is just to lower the temperature
further~\cite{Spallucci09}.   

\section{Results} \label{sec4}

Previous studies of black hole remnants considered a high initial mass
($\sim 10$~TeV) black hole and examined its decay down to the Planck
scale~\cite{Rizzo06b,Koch05}. 
This allowed for the study of the decay over a significant mass range.
However, black holes in proton-proton collisions are typically produced
just above their mass threshold, and large mass-range decay schemes
occur infrequently.  
The probability of such decays are reduced by a couple orders of
magnitude when compare to those near threshold.
In this study, we properly weight the production probability by the
parton distribution functions of the proton.

To preform the studies, we have modified the black hole Monte Carlo event 
generator CHARYBDIS~\cite{Harris03a,Frost09}, to use the noncommutative
gravitational radius in Eq.~(\ref{eq3}).
The CTEQ6L1 parton distribution functions~\cite{Pumplin} were used with
a QCD scale of $Q = 1/r_g$.
We set the minimum mass to be the Planck scale.  
If the remnant mass is above the Planck scale, we set the minimum mass to
be the remnant mass.
We thus produce black holes over the entire allowed mass range without
introducing any unphysical threshold to ensure that we are in a
semiclassical regime.
This is justified as noncommutative geometry inspired black holes can be
thought of as an effective theory for quantum black holes.
There will be some uncertainty near the Planck scale, but those black
holes with remnants above the Planck scale should be less effected by
these uncertainties. 

\FIGURE[h]{
\epsfig{file=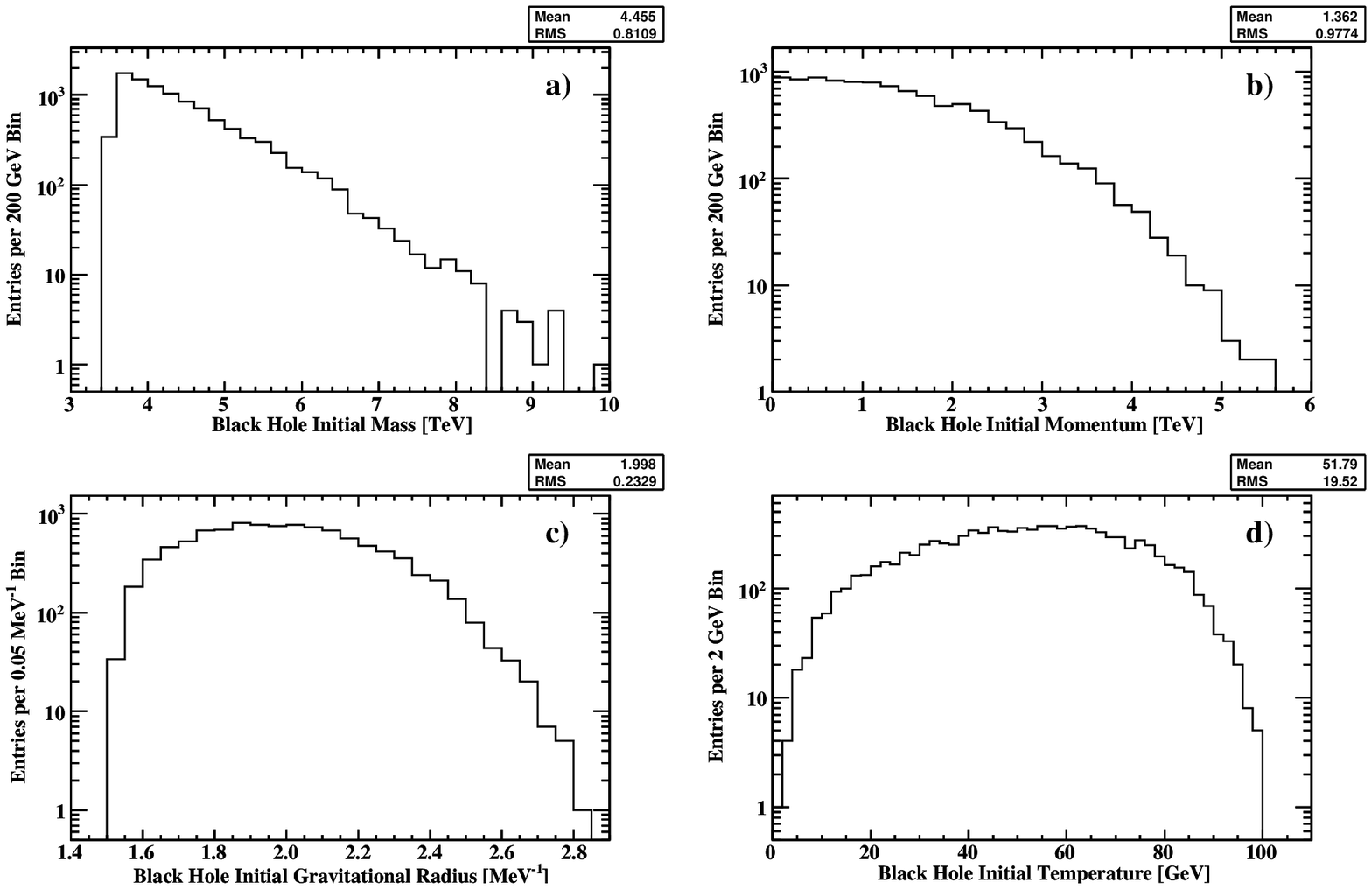,width=\columnwidth}
\caption{Noncommutative inspired black hole initial characteristics for 
$n = 4$,
$M_D = 0.94$~TeV, 
$\sqrt{\theta} = 0.64$~TeV$^{-1}$,
and a remnant mass of 3.6~TeV: a) mass, b) momentum, c) gravitational
radius, and d) temperature.} 
\label{fig4}
}

As an interesting representative case, we consider the $n = 4$
noncommutative inspired black hole with $\sqrt{\theta} = 0.64$ and
$M_D = 0.94$~TeV. 
This give a remnant mass of 3.6~TeV, about four times the Planck scale. 
The proton-proton cross section is about 1~nb for a centre-of-mass
energy of 14~TeV, which would result in significant production.
Figure~\ref{fig4} shows some of the initial black hole characteristics
for this case.
The most probable produced black hole mass is 3.6~TeV, but the mean mass
is about 0.9~TeV higher.  
The production probability has dropped by about two orders of magnitude
for an initial mass of about 7.5~TeV.
Thus high-mass black holes of $\mathcal{O}(10)$~TeV are produced
relatively infrequently, but with such large cross sections can still be
produced. 
We have not given the black hole any transverse momentum to the beam-line
at production.
Its initial momentum is taken to be entirely along the beam-line.
The most probable value of the initial black hole momentum is near zero,
but values can be as high as a few TeV.
The initial black hole gravitational radius is distributed over a fairly
narrow range. 
It is never more than about twice its minimum value.
The initial temperature is less than about 100~GeV, with a mean value of
about 50~GeV.
This justifies our statement about the noncommutative black hole being
cold, since the temperature is significantly below the minimum black
hole mass or the Planck scale.

For the decays, we employ Hawking evaporation using the modified
temperature in Eq.~(\ref{eq8}).  
So as not to add extraneous confusing effects, we have disabled spin and
greybody factors in CHARYBDIS. 
Graviton emission is not simulated.
The type of particle in each emission is determined by the normalized
flux spectra in the usual way in CHARYBDIS~\cite{Frost09}.
CHARYBDIS minimizes the charge and baryon number in each emission.
This is not necessary for charge, but removing the requirement has
little effect~\cite{Koch05}.
The power spectra are used to determine the decay particle energy.
If the resulting decay kinematics are violated, the generated particle is
rejected and another particle is generated.
Near the remnant mass, the kinematics are easily violated, so for
technical reasons we end the evaporation when the black hole mass is
within 100~MeV of its remnant mass.
The 100~MeV cutoff stops $u$, $d$, $g$, or $e$ production.
It also avoids low energy quarks that would make hadronization
difficult.
Unfortunately, the generator is still very inefficient when the black
hole is near the remnant mass.
If the remnant has nonzero baryon number or magnitude of electric charge
that is greater than unity, the entire decay is rejected and a new decay
is attempted.  
If after 200 attempts, the decay is not possible, new event kinematics
are chosen and another decay with the new kinematics is attempted.

\FIGURE[h]{
\epsfig{file=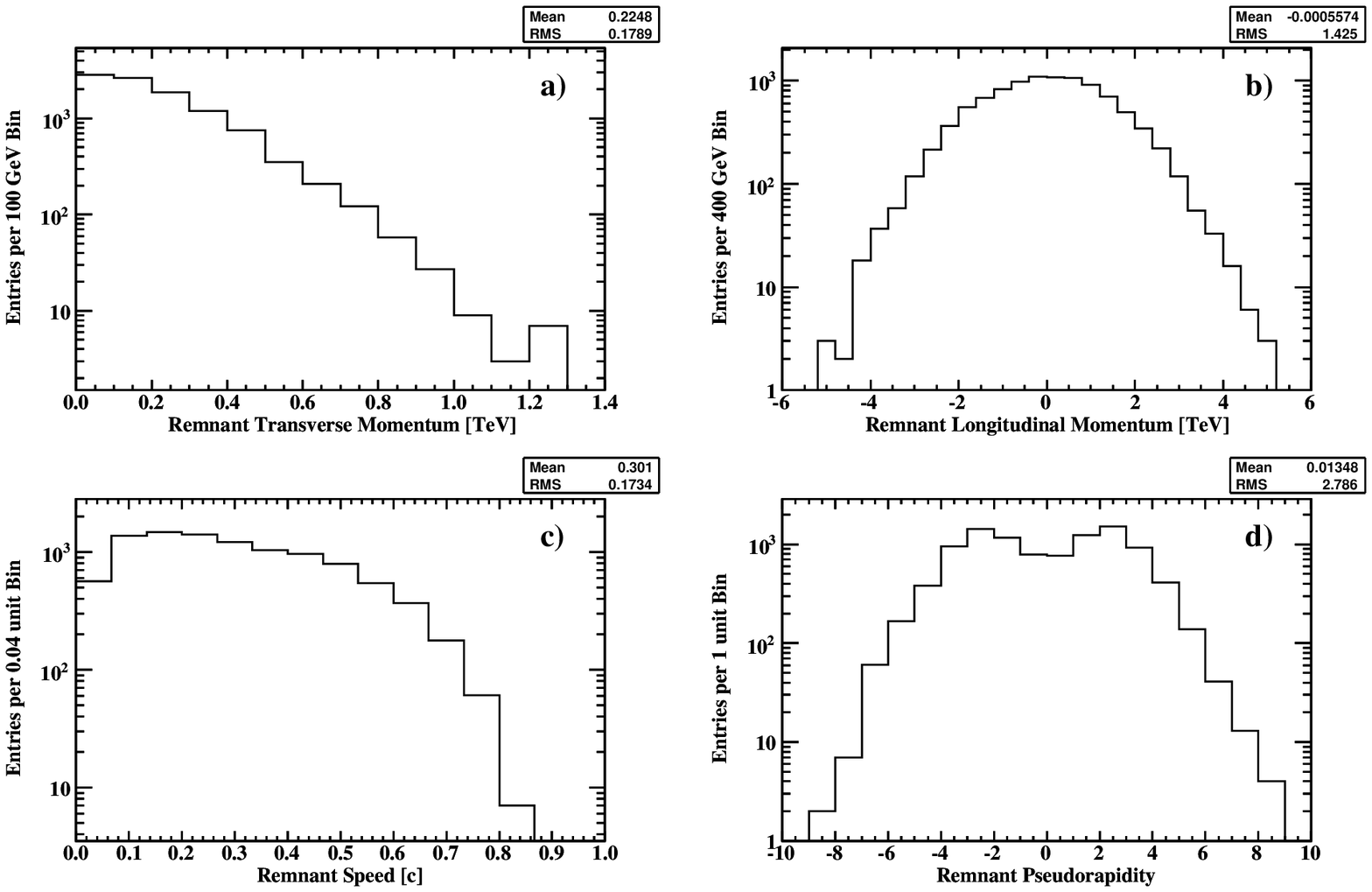,width=\columnwidth}
\caption{Noncommutative inspired black hole remnant characteristics for
$n=4$, 
$M_D = 0.94$~TeV, 
$\sqrt{\theta} = 0.64$~TeV$^{-1}$,
and a remnant mass of 3.6~TeV: a) transverse momentum, b) longitudinal
momentum, c) speed, and d) pseudorapidity.}
\label{fig5}
}

\FIGURE[h]{
\epsfig{file=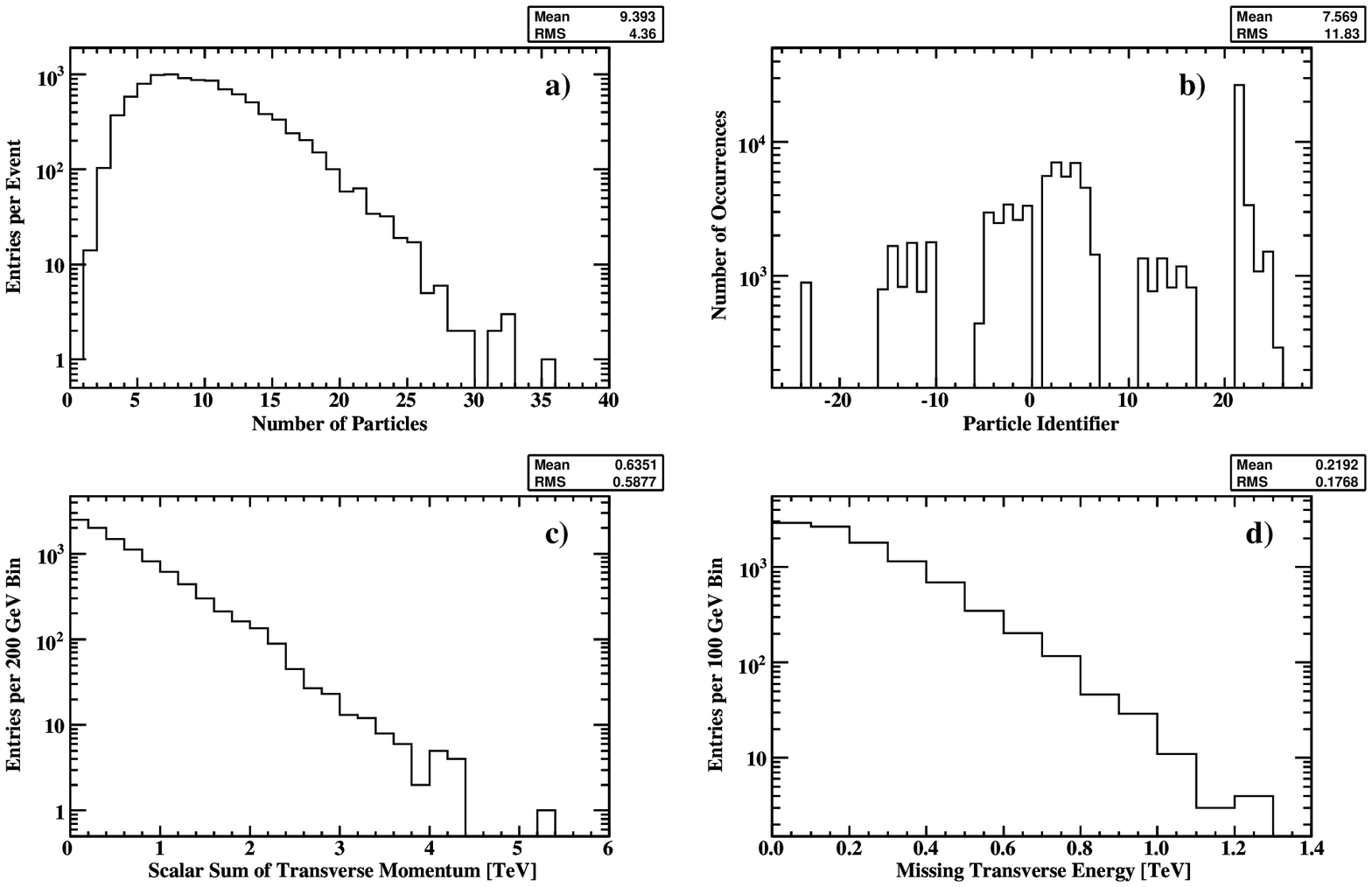,width=\columnwidth}
\caption{Noncommutative inspired black hole event characteristics for 
$n = 3$, 
$M_D = 0.94$~TeV, 
$\sqrt{\theta} = 0.64$~TeV$^{-1}$, 
and a remnant mass of 3.6~TeV: a) number of primary particles, b) PDG
particle identifier codes, c) scalar sum of transverse momentum, and d)
missing transverse energy.}
\label{fig6}
}

The remnant thus has zero baryon number and is either neutral or has
unit charge.
Vanishing baryon number allows the colour flow to be made for the
subsequent hadronization stage.
In this study, we should only consider neutral remnants since we have
not considered electric charge in the metric.
However, the decays are not significantly different for a charged remnant
and the probability of a charged remnant occurring in our simulation is
only about 3\%.
In any case, we do not examine the charge of the remnant.
Figure~\ref{fig5} shows the characteristics of the black hole remnant.
The remnant picks up very little transverse momentum, and is on average
less than 230~GeV. 
The longitudinal momentum is larger but the most probably value is still
small. 
Thus, the most probable speed is about $0.3c$ and normally less than
$0.8c$. 
About half of the remnants have a pseudorapidity $|\eta| > 2.5$ and will
thus miss the central detectors of ATLAS and CMS.
The higher the initial black hole mass, relative to its minimum mass, the
lower the pseudorapidity of the remnant.
In any case, if the remnant is neutral, it is unlikely to be detected.
In summary, the remnant is non-relativistic and has a fair chance of
going forward with vary little transverse momentum, when compared to its
mass. 
Hence, we expect the decay particles also to have low transverse momentum.

\FIGURE[h]{
\epsfig{file=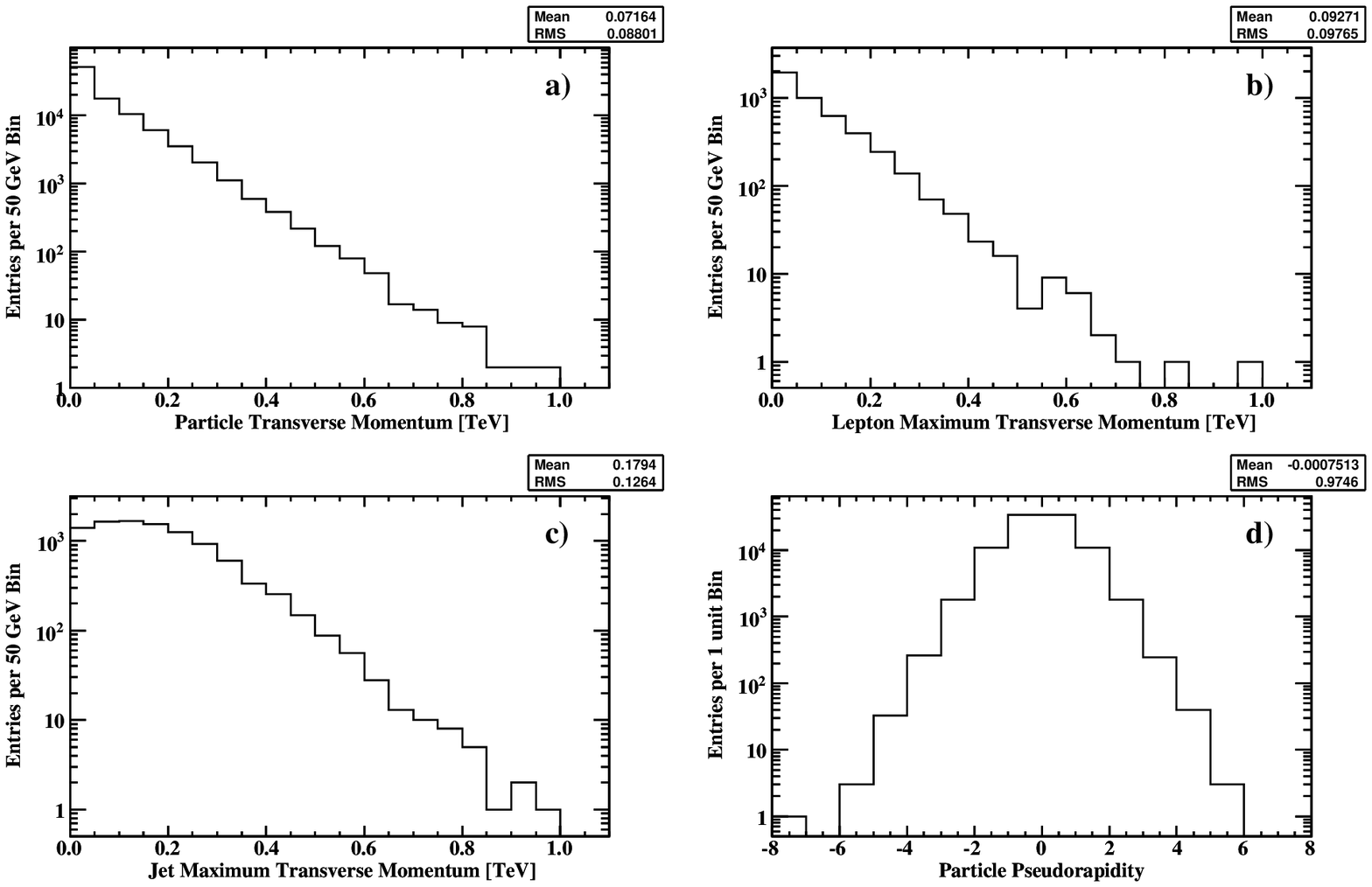,width=\columnwidth}
\caption{Noncommutative inspired black hole decay particle
characteristics for 
$n = 3$,
$M_D = 0.94$~TeV, 
$\sqrt{\theta} = 0.64$~TeV$^{-1}$, 
and a remnant mass of 3.6~TeV: primary particle transverse momentum, b)
lepton maximum transverse momentum, c) jet maximum transverse momentum,
and d) primary particle pseudorapidity.}
\label{fig7}
}

Figure~\ref{fig6} shows some of the event characteristics.
Although the transverse momentum is low, the mean number of primary
decay particles from the black hole is about 9 and can be as high as
about 25. 
The multiplicity is high due to the large number of emissions of low
energy particles in the later stage of decay.
It would be even higher if we had not imposed a technical cutoff above
the remnant mass. 
By examining the particle identifiers, we see the usual asymmetries from
total democracy because of the need to conserve baryon number and
charge~\cite{Gingrich07a}. 
Significant in this case is the reduced number of top quarks, which are 
suppressed due to the limited momentum phase space. 
Experiments searching for black holes will typically reconstruct the
scalar sum of the transverse momentum and missing transverse
energy~\cite{CSC,ATLAS-PHYS-PUB-2009-011,CMS_TDR}. 
The scalar transverse momentum is correlated with the parton-parton
centre-of-mass energy in $s$-channel processes.  
For low-mass semiclassical black hole searches at the LHC, a typical
requirement would be that the scalar sum of transverse momentum be above 
a few TeV~\cite{Gingrich08a}.  
Since the scalar sum of the transverse momentum extends down to
zero in the noncommutative case, imposing a selection on this variable
would significantly reduced the efficiency of searches. 
Ref.~\cite{Koch05} was the first to notice this effect.
Since the remnant is undetected it results in a huge missing energy.
However, only the missing transverse energy can be calculated in a
proton-proton collider. 
The remnant has relatively small transverse momentum and gives rise to a
mean missing transverse energy of about 200~GeV in the events.
This might be comparable to top-quark events and is not higher than
many other beyond the standard model physics scenarios. 

Some characteristics of the decay particles are shown in Fig.~\ref{fig7}.
The transverse momentum spectrum of the final state particles is soft,
with a mean of only about 70~GeV. 
It would be even softer if we had not imposed a technical cutoff above
the remnant mass. 
Most of the momentum is taken by the longitudinal momentum of the
remnant itself.
When searching for black holes in experiments, studies typically select
a few high transverse momentum $p_T$ jets, and perhaps reduce QCD
background by requiring a high transverse momentum lepton (electron or  
muon)~\cite{CSC,ATLAS-PHYS-PUB-2009-011,CMS_TDR}. 
For our example, most of the events contain at least one jet and about
45\% of the events contain a primary lepton.
Because of the rather soft maximum-$p_T$ jet and maximum-$p_T$ lepton
spectra, search strategies would have to be re-optimized.
This can only be done by the experiments, using a full detector
simulation.  
Although the particles have low transverse momentum, most of them have
pseudorapidity of $|\eta| < 2.5$ and should be detectable in the central 
detectors of ATLAS and CMS.

The decay final states are dominated by the characteristics of the
remnant. 
The signatures are very different from the high-multiplicity, high-$p_T$
states in semiclassical black holes and string balls, and very different 
from the high-$p_T$ two-body final states studied in gravitational
scattering, or presumably by quantum black holes produced near the Planck
scale~\cite{Gingrich09b}.
Because of the remnant, search strategies will have to be significantly
revised.
The signature will be some missing energy and a lot of soft particles.
The triggering strategies will have to be reexamined.
Because of the soft particles, it will be rather difficult to use
$p_T$-balance to get information on the remnant. 

Another possibility is to detect the remnant directly.
If the remnant is electrically neutral, it can only interact
gravitationally and will leave the detector.
Even if the remnant is charged, there is a fair chance it will go
forward.
We have only allowed the remnant to have unit charge, but in principle
the remnant could have a higher charge.
A more detailed treatment including charge effects should be performed.
It is also then important to consider the detector capabilities for
detecting massive charged particles.

Unless quark and gluon emission is suppressed as the black hole
evaporates down to the remnant, the remnant may also have colour charge.
How the remnant can shed its colour charge, yet not become lighter is
unknown.  
We have assumed the black hole sheds its colour during evaporation to  
become a colourless remnant.
We have not assumed the same for electric charge.

\section{Discussion} \label{sec5}

Mathematically, black holes can have any positive radius or mass in the
usual commutative ADD scenario.
It is commonly assumed that the threshold for black hole production is
the Planck scale. 
The shape of the threshold is usually taken as a step function.
However, it is anticipated that quantum and/or stringy corrections will
smooth out the threshold behaviour.
In addition, the threshold region will be more complicated due to
gravitational radiative corrections. 
To ensure the validity of the predictions of semiclassical general
relativity, a lower threshold is usually imposed on the black hole masses
that can be safely considered. 
This model-dependent threshold is often taken to be five times higher
than the Planck scale.
However, if semiclassical black holes can be produced, in real data
there should also be nonperturbative gravitational objects produced
below this threshold, but the models have no predictive power in this
mass region. 
This is the domain of quantum gravity and quantum black holes.

When searching for black holes at the LHC, Monte Carlo simulation
programs for the production and decay of black holes are used.
To avoid the non-predictive region of quantum gravity, the above
mentioned non-physical threshold is imposed in the simulations.
In most cases, this threshold has more of an impact on the anticipated
signatures than the physical parameters $M_D$ and $n$, or the decay
assumptions. 
The resulting cross section is an exponential or power-law looking
distribution in mass with a lower-mass threshold. 
Experimental effects will fold the mass distribution with a
gaussian-like resolution function, and the resulting mass distribution
will be a gaussian with a maximum approximately at the non-physical
threshold~\cite{CSC,ATLAS-PHYS-PUB-2009-011,CMS_TDR}.
But if low-scale gravity is realized, it is unlikely that nothing will be
produced below the model-dependent mass threshold.
The gaussian mass distribution will likely have a significantly
distorted low-mass tail due quantum black holes that pass the
experimental acceptance. 
In fact, a perfectly efficient search with no acceptance restrictions
would allow all the quantum black holes to be observed, and thus, shift
the maximum of the gaussian mass distribution downward towards the
Planck scale, by an amount depending on the interplay between the
quantum and semiclassical regimes.   
In this case, clearly, the Monte Carlo simulations do not accurately
simulate the data.
Noncommutative inspired black holes could offer a way out of this
difficulty by having a physical lower-mass threshold above the Planck
scale, and thus eliminating the need for a model-dependent threshold.
The benefits of this scenario would, of course, depend on the actually
fundamental constants that nature has chosen relative to the LHC energy
reach. 

Noncommutative gravity allows us to extend the validity of our models.
Once we leave the regime of applicability of semiclassical gravity, we
enter the regime of quantum gravity. 
In noncommutative theories, when we leave the regime of validity of
commutative semiclassical gravity, we enter the noncommutative
regime. 
However, noncommutative gravity can be only considered as an effective
theory, not a replacement for quantum gravity. 
Eventually, other quantum effects due to a UV complete theory of quantum
gravity will appear. 
Thus noncommutative gravity may allow a smooth transition in our
understanding between the semiclassical and strong quantum gravity
regime, thus extending the range of validity of predictions.

One of the aims of this paper is to guide experimentalists in searching
for possible signatures of quantum gravity.
Searches have been planned for semiclassical black holes well above the
Planck scale and for gravitational scattering below the Planck scale,
but no strategies exist for searches at or just above the Planck scale.  
String inspired search strategies have extended the semiclassical black
hole regime down to lower masses, but still above the mass scale for new
physics~\cite{ATLAS-PHYS-PUB-2009-011}.
Noncommutativity and extra dimensions, both motivated by
string/M-theory, allow the considerations of noncommutative geometry
inspired black holes.
This offers an alternative search strategy for the effects of quantum
gravity down to the Planck scale. 
Without such studies, as presented here, to guide experiments, there are
an intractable number of possible model independent searches that can be
form. 
However, the danger of phenomenological guidance is that searches can
become too narrowly focused.
By exploring alternative models, such as that considered here, it is
hoped that gravitational searches at the LHC will be inclusive enough
not to miss signatures for new physics.  

\acknowledgments

This work was supported in part by the Natural Sciences and Engineering
Research Council of Canada.

\bibliographystyle{JHEP}
\bibliography{gingrich}
\end{document}